# Quasi-Linear Transfer Function: A New Method for Frequency Domain Analysis of Nonlinear Systems

Hua-Liang Wei,   S.A. Billings
Department of Automatic Control and Systems Engineering, University of Sheffield
Mappin Street, Sheffield, S1 3JD, UK

**Abstract**: A new concept, called quasi-linear transfer functions (QLTF), which can be used to characterize the output frequency behaviour of nonlinear systems, is introduced based on the well-known Volterra series representation. By using the new concept of QLTF, it can be proved that the input and output frequency behaviour of a given system can be expressed using a number of one-dimensional functions with a form similar to that of the traditional frequency response function for linear systems. Two algorithms, which can be used to determine the valid range of the associated output frequencies of arbitrary order nonlinear subsystems with both a multitone and general inputs, are provided. The results obtained provide a new important insight into the output frequency characteristics of nonlinear systems and have many potential applications in nonlinear systems analysis and nonlinear structure detection.

**Keywords**:  Frequency domain analysis;  generalized frequency response functions;  nonlinear systems;  quasi-linear transfer function;  Volterra series

## 1. Introduction

In linear systems theory, the frequency response function or transfer function plays a key role in frequency domain analysis and provides a thorough insight into the system output behaviour. Frequency domain analysis therefore provides a powerful tool for linear system analysis, design and control. Motivated by the successful applications of the linear frequency response function and associated techniques, a similar concept, the generalised frequency response functions(GFRFs), were proposed in the late 1950s (Brilliant 1958, George 1959, Barrett 1963). Most of these results were based on the well known Volterra series theory, see for example, Wiener (1958), Volterra (1959), Bedrosian and Rice (1971), and Bussgang et al. (1974). Although it would be desirable to develop techniques for nonlinear system analysis in the frequency domain which are similar to the linear system, progress in this direction has been very slow due to the difficulty of estimating the frequency response functions of practical nonlinear systems, and the difficulty in the representation and interpretation of high-order nonlinear frequency response functions. Input-output representations based on the Volterra series theory were revived in the 1980s, partly due to the enhanced capability of modern computers for estimating high-order kernels and calculating high-order generalized nonlinear frequency response functions (GFRFs). This was helped by the appearance of several books and numerous papers which provided theoretical foundations of associated techniques, including the books by Marmarelis and Marmarelis (1978), Schetzen (1980a) and Rugh (1981), and numerous papers including Chua and Ng (1979a, 1979b), Masri and Caughey (1979), Billings (1980), Schetzen (1980b), Tomlinson (1980), Tang et al. (1983), Simon and Tomlinson (1984), Boyd and Chua (1985), Powers and Miksad (1987), Vinh et al. (1987), Billings and Tsang (1989a, 1989b, 1990), Peyton Jones and Billings (1989), and Gifford and Tomlinson (1989).

The computational problems associated with estimating the GFRFs directly from input-output observational data sets and the difficulties in physical interpretation of these functions have strongly limited the application of GFRFs. Recent developments relating to the connection between nonlinear spectral analysis (Billings and Tsang 1989a, 1989b, 1990) and the NARMAX methodology (Leontaritis and Billings 1985), has, however, overcome many of the difficulties associated with the estimation of GFRFs, which can now easily be obtained from time domain models, see for example, Peyton Jones and Billings (1989), Billings and Peyton Jones (1990), and Zhang and Billings (1993, 1994). By applying a recursive probing or harmonic generating algorithm (Peyton Jones and Billings 1989) to the identified NARMAX models or known nonlinear differential equations, it is now possible to obtain the nonlinear GFRFs for many real systems (Storer and Tomlinson 1993, Worden et al. 1994, Palumbo and Piroddi 2000, Boaghe et al. 2001). This has consequently renewed the interest in this approach and made the analysis of nonlinear GFRFs more applicable.

The GFRF-based approach for general nonlinear systems is different from the traditional linear approach in two principle aspects. Firstly, the frequency response function (FRF) representation of an equivalent time-domain nonlinear system consists of a finite or infinite number of FRFs instead of only one function as in the linear case. Secondly, each nonlinear FRF is always a multivariate function even when the underlying system is single-input and single-output (SISO). This not only increases the difficulty of analysis but also makes it very complicated to relate the FRF to the system response characteristics and to provide a physical interpretation of the dominant effects. A great effort has been made to interpret the GFRFs associated with nonlinear systems using both analytically quantitative analysis and graphically qualitative methods (Peyton and Billings 1990, Zhang and Billings 1993, Lang and Billings 1996, Billings and Lang 1996, Billings and Yusof 1996, Lang and Billings 2000). Most existing GFRF-based approaches attempt to provide analytical expressions or graphical interpretations which are useful for characterizing the system output behaviour in the frequency domain for some specific situations.

One important aspect of system analysis in the frequency domain is to investigate the relationship between the system input frequencies and the output frequency behaviour. For linear systems, the output frequency function $Y(j\omega)$ is related to the input frequency spectrum $U(j\omega)$ by the system frequency response function $H(j\omega)$ via the simple linear relationship $Y(j\omega) = H(j\omega)U(j\omega)$. This simple basic result provides the foundation for linear system analysis and design in the frequency domain. In this case, the input frequencies pass independently through the system, that is, an input at a given frequency $\omega$ produces at steady state an output at the same frequency and no energy is transferred to or from any other frequency components. The system frequency response function $H(j\omega)$ itself alone can totally characterise a given linear system. Inspired by this basic result for linear systems, the central objective of this study is to introduce a similar concept to describe the relationship between the input and output frequency spectra of nonlinear systems using GFRFs. A new concept, called *quasi-linear transfer function* (QLTF), is introduced as one-dimensional functions with a form similar to the traditional frequency response function for linear systems. The introduction of QLTF will not only yield a much simpler description for the output spectrum but also provides a totally new viewpoint on the decomposition of nonlinear system output frequency responses. This provides great potential for nonlinear system analysis in the frequency domain. Because the range of output frequencies are different from that of the input for a nonlinear system, new algorithms for determining the output frequency range of arbitrary order nonlinear outputs are also derived.

## 2. Generalised frequency response functions for nonlinear systems

It is well known that the input-output relationship of a wide class of nonlinear systems can be approximated in the time domain by the Volterra functional series (Schetzen 1980a, Billings and Tsang 1989a, 1989b, Peyton Jones and Billings 1990)

$$y(t) = \sum_{n=1}^{N} y_n(t) \tag{1}$$

where the system output $y(t)$ is expressed as a sum of the response of a sequence of parallel subsystems, each of which is related to both the system input $u(t)$ and an $n$th-order kernel. The output of the $n$th-order nonlinear subsystem, $y_n(t)$, is characterised by an extension of the familiar convolution integral of linear system theory to higher dimensions

$$\begin{aligned} y_n(t) &= \int_{-\infty}^{\infty} \cdots \int_{-\infty}^{\infty} h_n(\tau_1, \cdots, \tau_n) u(t-\tau_1) \cdots u(t-\tau_n) d\tau_1 \cdots d\tau_n \\ &= \int_{-\infty}^{\infty} \cdots \int_{-\infty}^{\infty} h_n(\tau_1, \cdots, \tau_n) \prod_{i=1}^{n} [u(t-\tau_i) d\tau_i] \end{aligned} \tag{2}$$

where the *nth-order kernel* or *nth-order impulse response* $h_n(\tau_1, \cdots, \tau_n)$ is so called because this reduces to the linear impulse response function for the simplest case $n=1$.

In order to obtain an equivalent description of (2) in the frequency domain, it is common to introduce the so called *nth-order associated function*

$$y_n(t_1, \cdots, t_n) = \int_{-\infty}^{\infty} \cdots \int_{-\infty}^{\infty} h_n(\tau_1, \cdots, \tau_n) \prod_{i=1}^{n} [u(t_i - \tau_i) d\tau_i] \tag{3}$$

from which the desired output $y_n(t)$ can be recovered by the restriction

$$y_n(t) = y_n(t_1, \cdots, t_n)\big|_{t_1 = \cdots = t_n = t} \tag{4}$$

Apply a multidimensional Fourier transform to both sides of Eq. (3), yields

$$Y_n(j\omega_1, \cdots, j\omega_n) = H_n(j\omega_1, \cdots, j\omega_n) \prod_{i=1}^{n} U(j\omega_i) \tag{5}$$

where $U(\cdot)$ is the input spectrum defined as $U(j\omega) = \mathbf{FT}[u(t)]$, where $\mathbf{FT}[\cdot]$ denotes the Fourier transform operator. $H_n(j\omega_1, \cdots, j\omega_n)$ is the *nth-order transfer function* or *nth-order generalised frequency response function* (GFRFs) defined as

$$H_n(j\omega_1, \cdots, j\omega_n) = \int_{-\infty}^{\infty} \cdots \int_{-\infty}^{\infty} h_n(\tau_1, \cdots, \tau_n) e^{-j(\omega_1 \tau_1 + \cdots + \omega_n \tau_n)} d\tau_1 \cdots d\tau_n \tag{6}$$

Note that Eq. (6) reduces to the standard linear transfer function for the case of $n=1$, and that Eq. (5) would then give the familiar one-dimensional linear relationship

$$Y_1(j\omega) = H_1(j\omega)U(j\omega) \tag{7}$$

A basic property of linear systems, illustrated by the relation (7), is that input frequencies pass independently through the system. Thus an input at a given frequency $\omega$ produces at steady state an output at the same frequency and no energy is transferred to or from any other frequency components.

From Eqs. (3), (4) and (5), it is clear that

$$\begin{aligned} y_n(t) &= \mathbf{FT}^{-1}[Y_n(j\omega_1,\cdots,j\omega_n)] \\ &= \frac{1}{(2\pi)^n}\int_{-\infty}^{\infty}\cdots\int_{-\infty}^{\infty}Y_n(j\omega_1,\cdots,j\omega_n)e^{j(\omega_1+\cdots+\omega_n)t}d\omega_1\cdots d\omega_n \end{aligned} \tag{8}$$

where $\mathbf{FT}^{-1}[\cdot]$ is used to denote the inverse Fourier transform. In fact the relationship (8) can be derived directly from Eq. (2) without the help of the $n$th-order associated function (3) (Lang and Billings 1996). This is shown below.

$$\begin{aligned} y_n(t) &= \int_{-\infty}^{\infty}\cdots\int_{-\infty}^{\infty} h_n(\tau_1,\cdots,\tau_n)u(t-\tau_1)\cdots u(t-\tau_n)d\tau_1\cdots d\tau_n \\ &= \int_{-\infty}^{\infty}\cdots\int_{-\infty}^{\infty} h_n(\tau_1,\cdots,\tau_n)\prod_{i=1}^{n}\left[\frac{1}{2\pi}\int_{-\infty}^{\infty}U(j\omega_i)e^{j\omega_i(t-\tau_i)}d\omega_i\right]d\tau_1\cdots d\tau_n \\ &= \frac{1}{(2\pi)^n}\int_{-\infty}^{\infty}\cdots\int_{-\infty}^{\infty}\left[\int_{-\infty}^{\infty}\cdots\int_{-\infty}^{\infty}h_n(\tau_1,\cdots,\tau_n)\prod_{i=1}^{n}[e^{-j\omega_i\tau_i}d\tau_i]\right]\prod_{i=1}^{n}\left[U(j\omega_i)e^{j\omega_i t}d\omega_i\right] \\ &= \frac{1}{(2\pi)^n}\int_{-\infty}^{\infty}\cdots\int_{-\infty}^{\infty}H_n(j\omega_1,\cdots,j\omega_n)\prod_{i=1}^{n}\left[U(j\omega_i)e^{j\omega_i t}d\omega_i\right] \\ &= \frac{1}{(2\pi)^n}\int_{-\infty}^{\infty}\cdots\int_{-\infty}^{\infty}Y_n(j\omega_1,\cdots,j\omega_n)e^{j(\omega_1+\cdots+\omega_n)t}d\omega_1\cdots d\omega_n \end{aligned} \tag{9}$$

The meaning and characteristics of the multidimensional frequency function $Y_n(j\omega_1,\cdots,j\omega_n)$ have been discussed in detail by Peyton Jones and Billings (1990) and Lang and Billings (1996).

## 3. The concept of quasi-linear transfer function

### 3.1 Output frequency functions in the input/output frequency domain

Consider firstly the relationship between the system input spectrum and the output frequency behaviour. For linear systems, the output frequency function $Y(j\omega)$ is connected to the input frequency function $U(j\omega)$ by the system frequency response function $H(j\omega)$ via a simple linear relationship

$$Y(j\omega) = H(j\omega)U(j\omega) \tag{10}$$

This simple basic result provides the foundation for linear system analysis and design in the frequency domain. In this case, the system frequency response function $H(j\omega)$ itself alone can totally characterise a given linear system. For nonlinear systems, however, similar result is not available. This study attempts to build such a relationship for nonlinear systems by introducing a new concept of quasi-linear transfer function (QLTF), which will be described below.

By making a change of variables

$$\begin{cases} \sigma_i = \omega_i, \ 1 \leq i \leq n-1 \\ \sigma_n = \sum_{i=1}^{n} \omega_i \end{cases} \tag{11}$$

Eq. (9) becomes

$$y_n(t) = \frac{1}{(2\pi)^n} \int_{-\infty}^{\infty} \cdots \int_{-\infty}^{\infty} Y_n(j\sigma_1, \cdots, j\sigma_{n-1}, j(\sigma_n - \sigma_1 - \cdots - \sigma_{n-1})) e^{j\sigma_n t} d\sigma_1 \cdots d\sigma_n \tag{12}$$

$$= \frac{1}{2\pi} \int_{-\infty}^{\infty} \left[ \frac{1}{(2\pi)^{n-1}} \underbrace{\int_{-\infty}^{\infty} \cdots \int_{-\infty}^{\infty}}_{n-1} Y_n(j\omega_1, \cdots, j\omega_{n-1}, j(\omega - \sum_{i=1}^{n-1} \omega_i)) d\omega_1 \cdots d\omega_{n-1} \right] e^{j\omega t} d\omega$$

$$= \frac{1}{2\pi} \int_{-\infty}^{\infty} Y_n(j\omega) e^{j\omega t} d\omega \tag{13}$$

where

$$Y_n(j\omega) = \frac{1}{(2\pi)^{n-1}} \int_{-\infty}^{\infty} \cdots \int_{-\infty}^{\infty} Y_n(j\omega_1, \cdots, j\omega_{n-1}, j(\omega - \omega_1 - \cdots - \omega_{n-1})) d\omega_1 \cdots d\omega_{n-1} \tag{14}$$

In Eq (14), the family $\{\omega_1, \cdots, \omega_n; \omega\}$ was referred to as the *input/output frequency domain* in Peyton Jones and Billings (1990). $Y_n(j\omega)$ can therefore be referred to as the *nth-order input/output frequency function*. For a physical interpretation of Eqs. (9) and (14), see the papers by Peyton Jones and Billings (1990) and Lang and Billings (1996). Note from the variable transform (11) that the input/output frequency domain is restricted to $\omega_1 + \cdots + \omega_n = \omega$. The valid frequency range of the output spectrum can therefore be determined by the input frequencies. Denote the valid frequency domain of the $n$th-order input-output frequency function $Y_n(j\omega)$ as $\Omega_n$ ($n = 1, 2, \cdots, N$). $Y_n(j\omega)$ is thus defined on $\Omega_n$. The definition of $\Omega_n$ will be discussed later in detail.

3.2 Quasi-linear transfer function

It can be seen by observing Eq. (2) that the $n$th-order nonlinear subsystem is excited by the $n$th order input, that is, by $u_n(t) = u(t - \tau_1) u(t - \tau_2) \cdots u(t - \tau_n)$. From the time-shift property of the Fourier transform, the amplitude spectrum of $u_n(t)$ is identical to that of $u^n(t)$, but each frequency component of the spectrum of $u_n(t)$ is shifted from that of $u^n(t)$ by an amount proportional to its frequency $\omega$.

In order to define the quasi-linear transfer functions, the *nth-order input spectral function* is introduced as below:

$$U_n(j\omega) = \mathbf{FT}[u^n(t)], \ n = 1, 2, \cdots, N \tag{15}$$

From the convolution theory in the frequency domain which states that

$$\mathbf{FT}[u(t)v(t)]_\omega = \frac{1}{2\pi}\mathbf{FT}[u(t)] * \mathbf{FT}[v(t)]_\omega = \frac{1}{2\pi}U(j\omega) * V(j\omega)$$

$$= \frac{1}{2\pi}\int_{-\infty}^{\infty} U(j\omega_1)V(j(\omega-\omega_1))d\omega_1 \tag{16}$$

It can be easily shown that

$$U_1(j\omega) = U(j\omega) \tag{17}$$

$$U_n(j\omega) = \frac{1}{(2\pi)^{n-1}} \int_{-\infty}^{\infty} \cdots \int_{-\infty}^{\infty} U(j\omega_1) \cdots U(j\omega_{n-1}) U(j(\omega - \omega_1 - \cdots - \omega_{n-1})) d\omega_1 \cdots d\omega_{n-1} \tag{18}$$

**Definition 3.1** *The nth-order quasi-linear transfer function (QLTF) of the nth-order nonlinear subsystem described by Eq. (2) is defined as*

$$G_n(j\omega) = \frac{Y_n(j\omega)}{U_n(j\omega)}, \quad \omega \in \Omega_n \tag{19}$$

*where $Y_n(j\omega)$ and $U_n(j\omega)$ are defined by (5) and (18) respectively. $\Omega_n$ is the valid frequency domain of $Y_n(j\omega)$.*

The quasi-linear transfer function $G_n(j\omega)$ is so called since the linear relationship

$$Y_n(j\omega) = G_n(j\omega)U_n(j\omega) \tag{20}$$

holds for $\omega \in \Omega_n$, $n=1,2,\cdots,N$. This is similar to that of the classical transfer function defined for linear systems. From (1), (13) and (20),

$$y(t) = \sum_{n=1}^{N} \frac{1}{2\pi} \int_{-\infty}^{\infty} Y_n(j\omega)e^{j\omega t} d\omega = \frac{1}{2\pi} \int_{-\infty}^{\infty} \left[\sum_{n=1}^{N} Y_n(j\omega)\right] e^{j\omega t} d\omega \tag{21}$$

Therefore, the system output frequency response or output spectrum to a given general input is

$$Y(j\omega) = \sum_{n=1}^{N} Y_n(j\omega) = \sum_{n=1}^{N} G_n(j\omega)U_n(j\omega), \quad \omega \in \bigcup_{n=1}^{N} \Omega_n \tag{22}$$

3.3   Properties of the QLTF

Two main properties of the QLTF are summarised below.

**Property 3.1**   *The nth-order quasi-linear transfer function $G_n(j\omega)$ is not affected by the amplitude of the input signal u(t).*

This is clear by observing that $G_n(j\omega)\big|_{cu(t)} = G_n(j\omega)\big|_{u(t)}$, where c is any real constant number.

**Property 3.2** *The nth-order quasi-linear transfer function $G_n(j\omega)$ is unique once the nth-order generalised frequency response function $H_n(j\omega_1,\cdots,j\omega_n)$ is determined.*

For a given nonlinear system, both the *n*th-order kernel $h_n(\tau_1,\cdots,\tau_n)$ and frequency response function $H_n(j\omega_1,\cdots,j\omega_n)$ may not be unique since changing the order of arguments may give different functions but will still yield the same output $y_n(t)$ and $Y_n(j\omega)$. In order to guarantee the uniqueness of $H_n(j\omega_1,\cdots,j\omega_n)$, a common practice is to symmetrise the function by summing all the asymmetric functions over all the permutations of the arguments and dividing by the number of the total asymmetric functions (Schetzen 1980a). Once the *n*th-order frequency response function $H_n(j\omega_1,\cdots,j\omega_n)$ is determined, the *n*th-order QLTF is then determined for a given input.

**Property 3.3** *The amplitude spectrum of the nth-order quasi-linear transfer function $G_n(j\omega)$ is symmetric about the origin, that is, $|G_n(-j\omega)|=|G_n(j\omega)|$ for any $\omega \in \Omega_n$, where $\Omega_n$ is the valid frequency domain of $G_n(j\omega)$.*

This property is easy to show by observing that from Eq. (18)

$$U_n^*(j\omega) = \frac{1}{(2\pi)^{n-1}} \int_{-\infty}^{\infty} \cdots \int_{-\infty}^{\infty} U^*(j\omega_1)\cdots U^*(j\omega_{n-1}) U^*(j(\omega-\omega_1-\cdots-\omega_{n-1}))d\omega_1\cdots d\omega_{n-1}$$

$$= \frac{1}{(2\pi)^{n-1}} \int_{-\infty}^{\infty} \cdots \int_{-\infty}^{\infty} U(-j\omega_1)\cdots U(-j\omega_{n-1}) U(j(-\omega+\omega_1+\cdots+\omega_{n-1}))d\omega_1\cdots d\omega_{n-1} \quad (23)$$

where the right-corner star "*" indicates the conjugate of the associated complex number. By making a change of variables as $\omega_i = -\sigma_i$ for *i*=1,2, …, *n*-1, Eq. (23) becomes

$$U_n^*(j\omega) = \frac{1}{(2\pi)^{n-1}} \int_{-\infty}^{\infty} \cdots \int_{-\infty}^{\infty} U(j\sigma_1)\cdots U(j\sigma_{n-1}) U(j(-\omega-\sigma_1-\cdots-\sigma_{n-1}))(-1)^{n-1} d\sigma_1\cdots d\sigma_{n-1}$$

$$= (-1)^{n-1} U_n(-j\omega) \quad (24)$$

Similarly,

$$Y_n^*(j\omega) = (-1)^{n-1} Y_n(-j\omega) \quad (25)$$

Therefore,

$$G_n^*(j\omega) = G_n(-j\omega) \quad (26)$$

and

$$|G_n(-j\omega)| = |G_n(j\omega)| \quad (27)$$

In Section 7 it will be shown using an example that the argument (or phase) spectrum of the *n*th-order quasi-linear transfer function $G_n(j\omega)$ is anti-symmetric about the origin for a multitone input discussed in the next section.

## 4. Derivation of the quasi-linear transfer function under a multitone input

The $n$th-order QLTF of a nonlinear system with a multitone input is a special case of the general input case which is described in Section 3. For the case of a multitone input, the $n$th-order QLTF can be greatly simplified and this is described below.

Consider the multitone input

$$u(t) = \sum_{i=1}^{K} |A_i| \cos(\omega_i t + \angle A_i) = \sum_{i=-K}^{K} \frac{A_i}{2} e^{j\omega_i t} \tag{28}$$

where $\omega_0 = 0$, $\omega_{-i} = -\omega_i$, $A_0 = 0$, $A_{-i} = A_i^*$. Noting that

$$U(j\omega) = \int_{-\infty}^{\infty} u(t) e^{-j\omega t} dt = \int_{-\infty}^{\infty} \sum_{i=-K}^{K} \frac{A_i}{2} e^{-j(\omega-\omega_i)t} dt$$

$$= \sum_{i=-K}^{K} \frac{A_i}{2} \int_{-\infty}^{\infty} e^{-j(\omega-\omega_i)t} dt = \sum_{i=-K}^{K} \pi A_i \delta(\omega - \omega_i) \tag{29}$$

where $\delta(\cdot)$ is the Kronecker delta function. The nth-order input spectral function $U_n(j\omega)$ defined by (18) is calculated to be

$$U_n(j\omega) = \frac{1}{(2\pi)^{n-1}} \int_{-\infty}^{\infty} \cdots \int_{-\infty}^{\infty} \left[ \prod_{i=1}^{n-1} \sum_{k_i=-K}^{K} \pi A_{k_i} \delta(\omega_i - \omega_{k_i}) \right] \times \sum_{k_n=-K}^{K} \pi A_{k_n} \delta(\omega - \sum_{i=1}^{n-1} \omega_i) d\omega_1 \cdots d\omega_{n-1}$$

$$= \frac{1}{2^{n-1}} \sum_{k_1=-K}^{K} \cdots \sum_{k_n=-K}^{K} \pi A_{k_1} \cdots A_{k_n}$$

$$\times \int_{-\infty}^{\infty} \cdots \int_{-\infty}^{\infty} \delta(\omega_1 - \omega_{k_1}) \cdots \delta(\omega_{n-1} - \omega_{k_{n-1}}) \delta(\omega - \sum_{i=1}^{n-1} \omega_{k_i}) d\omega_1 \cdots d\omega_{n-1}$$

$$= \frac{1}{2^{n-1}} \sum_{k_1=-K}^{K} \cdots \sum_{k_n=-K}^{K} \pi A_{k_1} \cdots A_{k_n}$$

$$\times \int_{-\infty}^{\infty} \cdots \int_{-\infty}^{\infty} \delta(\omega_1 - \omega_{k_1}) \cdots \delta(\omega_{n-1} - \omega_{k_{n-1}}) \delta(\omega - \sum_{i=1}^{n-1} \omega_{k_i}) d\omega_1 \cdots d\omega_{n-1}$$

$$= \frac{1}{2^{n-1}} \underbrace{\sum_{k_1=-K}^{K} \cdots \sum_{k_n=-K}^{K}}_{\omega_{k_1} + \cdots + \omega_{k_n} = \omega} \pi B(\omega_{k_1}) \cdots B(\omega_{k_n})$$

$$= \frac{1}{2^{n-1}} \sum_{k_1=-K}^{K} \cdots \sum_{k_{n-1}=-K}^{K} \pi B(\omega_{k_1}) \cdots B(\omega_{k_{n-1}}) B(\omega - \omega_{k_1} - \cdots - \omega_{k_n}) \tag{30}$$

where

$$B(\omega) = \begin{cases} A_k & \omega \in \{\omega_k : k = \pm 1, \cdots, \pm K\} \\ 0 & \text{otherwise} \end{cases} \tag{31}$$

Similarly, the $n$th-order input-output frequency function $Y_n(j\omega)$ is

$$Y_n(j\omega) = \frac{1}{(2\pi)^{n-1}} \underbrace{\int_{-\infty}^{\infty} \cdots \int_{-\infty}^{\infty}}_{n-1} Y_n(j\omega_1, \cdots, j\omega_{n-1}, j(\omega - \omega_1 - \cdots - \omega_{n-1})) d\omega_1 \cdots d\omega_{n-1}$$

$$= \frac{1}{2^{n-1}} \underbrace{\sum_{k_1=-K}^{K} \cdots \sum_{k_n=-K}^{K}}_{\omega_{k_1} + \cdots + \omega_{k_n} = \omega} \pi B(\omega_{k_1}) \cdots B(\omega_{k_n}) H_n(j\omega_{k_1}, \cdots, j\omega_{k_{n-1}}, j\omega_{k_n})$$

$$= \frac{1}{2^{n-1}} \sum_{k_1=-K}^{K} \cdots \sum_{k_{n-1}=-K}^{K} \pi B(\omega_{k_1}) \cdots B(\omega_{k_{n-1}}) B(\omega - \omega_{k_1} - \cdots - \omega_{k_{n-1}})$$

$$\times H_n(j\omega_{k_1}, \cdots, j\omega_{k_{n-1}}, j(\omega - \omega_{k_1} - \cdots - \omega_{k_{n-1}})) \qquad (32)$$

The $n$th-order QLTF can then be computed using Eqs. (30), (31) and (32).

## 5. QLTFs for discrete-time nonlinear systems

The concept of QLTF for the continuous-time models described in Sections 2 and 3 can be extended to the discrete-time case, where a nonlinear system is described using a discrete-time Volterra series as

$$y[k] = \sum_{n=1}^{N} y_n[k] \qquad (33)$$

where

$$y_n[k] = \sum_{i_1=-\infty}^{\infty} \cdots \sum_{i_n=-\infty}^{\infty} h_n[i_1, \cdots, i_n] u[k-i_1] \cdots u[k-i_n] \qquad (34)$$

and $h_n[i_1, \cdots, i_n]$ is the *nth-order discrete-time kernel* or *nth-order impulse response*. The symbol $x[\cdot]$ is adopted to indicate a discrete-time signal drawn from a continuous-time signal $x(\cdot)$ using equal sampling intervals $T$ in the sense that $x[k] = \{x(kT)\}_{k \in Z}$.

From Fourier transform theory, for any general sequence $x[k]$ that is of finite duration, the relations

$$X(j\omega T) = \sum_{k=-\infty}^{\infty} x[k] e^{-jk\omega T} \qquad (35)$$

$$x[k] = \frac{T}{2\pi} \int_{-\pi/T}^{\pi/T} X(j\omega T) e^{-jk\omega T} d\omega \qquad (36)$$

between the time-frequency domain sequence $x[k]$ and $X(j\omega T)$ define the *discrete-time Fourier transform* (DTFT) or semi-discrete Fourier transform pair, and $X(j\omega T)$ is referred to as the discrete-time Fourier transform of $x[\cdot]$. Note that the integration in (36) can be performed in any interval with a period of $2\pi/T$. Note that in practice it is common to make the change of variables $\Omega = \omega T$ and this brings about the equivalent definition for the discrete-time Fourier transform pair (35) and (36) as follows:

$$X(j\Omega) = \sum_{k=-\infty}^{\infty} x[k] e^{-jk\Omega} \qquad (35a)$$

$$x[k] = \frac{1}{2\pi}\int_{-\pi}^{\pi} X(j\Omega)e^{-jk\Omega}d\Omega \tag{36a}$$

The discrete-frequency variable $\Omega$ is related to the real-frequency variable $\omega$ by the equation $\Omega = \omega T$. The discrete frequency $\Omega$ can therefore be viewed as a scaled version of the real frequency $\omega$. Expanding Eq. (34) using the DTFT pair (35a) and (36a) yields

$$\begin{aligned}
y_n[k] &= \sum_{i_1=-\infty}^{\infty}\cdots\sum_{i_n=-\infty}^{\infty} h_n[i_1,\cdots,i_n]\prod_{m=1}^{n}\left\{\frac{1}{2\pi}\int_{-\pi}^{\pi} U(j\Omega_m)e^{j(k-i_m)\Omega_m}d\Omega_m\right\} \\
&= \left(\frac{1}{2\pi}\right)^n \int_{-\pi}^{\pi}\cdots\int_{-\pi}^{\pi}\left\{\sum_{i_1=-\infty}^{\infty}\cdots\sum_{i_n=-\infty}^{\infty} h_n[i_1,\cdots,i_n]e^{-j(i_1\Omega_1+\cdots+i_n\Omega_n)}\right\}\prod_{m=1}^{n}\{U(j\Omega_m)e^{jk\Omega_m}d\Omega_m\} \\
&= \left(\frac{1}{2\pi}\right)^n \int_{-\pi}^{\pi}\cdots\int_{-\pi}^{\pi} H_n(j\Omega_1,\cdots,j\Omega_n)\prod_{m=1}^{n}\{U(j\Omega_m)e^{jk\Omega_m}d\Omega_m\} \\
&= \left(\frac{1}{2\pi}\right)^n \int_{-\pi}^{\pi}\cdots\int_{-\pi}^{\pi} Y_n(j\Omega_1,\cdots,j\Omega_n)e^{jk(\Omega_1+\cdots+\Omega_n)}d\Omega_1\cdots d\Omega_n \\
&= \frac{1}{2\pi}\int_{-\pi}^{\pi}\left\{\left(\frac{1}{2\pi}\right)^{n-1}\underbrace{\int_{-\pi}^{\pi}\cdots\int_{-\pi}^{\pi}}_{n-1} Y_n(j\Omega_1,\cdots,j\Omega_{n-1},j(\Omega-\sum_{i=1}^{n-1}\Omega_i))d\Omega_1\cdots d\Omega_{n-1}\right\}e^{jk\Omega}d\Omega \\
&= \frac{1}{2\pi}\int_{-\pi}^{\pi} Y_n(j\Omega)e^{jk\Omega}d\Omega \tag{37}
\end{aligned}$$

where $Y_n(j\Omega_1,\cdots,j\Omega_n)$ is defined similar to (5), and

$$H_n(j\Omega_1,\cdots,j\Omega_n) = \sum_{i_1=-\infty}^{\infty}\cdots\sum_{i_2=-\infty}^{\infty} h_n[i_1,\cdots,i_n]e^{-j(i_1\Omega_1+\cdots+i_1\Omega_1)} \tag{38}$$

$$Y_n(j\Omega) = \left(\frac{1}{2\pi}\right)^{n-1}\int_{-\pi}^{\pi}\cdots\int_{-\pi}^{\pi} Y_n(j\Omega_1,\cdots,j\Omega_{n-1},j(\Omega-\Omega_1-\cdots-\Omega_{n-1}))d\Omega_1\cdots d\Omega_{n-1} \tag{39}$$

Denote

$$U_n(j\Omega) = \left(\frac{1}{2\pi}\right)^{n-1}\int_{-\pi}^{\pi}\cdots\int_{-\pi}^{\pi} U_n(j\Omega_1)\cdots,U(j\Omega_{n-1})U(j(\Omega-\sum_{i=1}^{n-1}\Omega_i))d\Omega_1\cdots d\Omega_{n-1} \tag{40}$$

The *n*th-order QLTF for a nonlinear system described by a discrete-time Volterra model can be defined as

$$G_n(j\Omega) = \frac{Y_n(j\Omega)}{U_n(j\Omega)}, \quad \Omega \in D_n \tag{41}$$

where $D_n$ is the valid frequency range of $Y_n(j\Omega)$.

For convenience of calculation, a simplified derivation of the DTFT pair (31) and (36) are often considered. Assume that $x[k]$ is a finite duration sequence of $N$ samples. From discrete Fourier transform theory, it can be proved that the relations

$$X[m] = \sum_{k=0}^{N-1} x[k] e^{-j2\pi mk/N} \quad (42)$$

$$x[k] = \frac{1}{N} \sum_{m=0}^{N-1} X[m] e^{j2\pi mk/N} \quad (43)$$

form a *discrete Fourier transform* (DFT) pair, and $X[\cdot]$ is referred to as the discrete Fourier transform of $x[\cdot]$. Expanding Eq. (34) using the DFT pair (42) and (43) yields

$$\begin{aligned}
y_n[k] &= \sum_{i_1=0}^{N-1} \cdots \sum_{i_n=0}^{N-1} h_n[i_1,\cdots,i_n] \prod_{\ell=1}^{n} \left\{ \frac{1}{N} \sum_{m_\ell=0}^{N-1} U[m_i] e^{j2\pi m_\ell(k-i_\ell)/N} \right\} \\
&= \frac{1}{N^n} \sum_{m_1=0}^{N-1} \cdots \sum_{m_n=0}^{N-1} \left\{ \sum_{i_1=0}^{N-1} \cdots \sum_{i_n}^{N-1} h_n[i_1,\cdots,i_n] e^{-j2\pi(m_1 i_1+\cdots+m_n i_n)/N} \right\} \\
&\qquad \times U[m_1]\cdots U[m_n] e^{j2\pi(m_1+\cdots+m_n)k/N} \\
&= \frac{1}{N^n} \sum_{m_1=0}^{N-1} \cdots \sum_{m_n=0}^{N-1} H_n[m_1,\cdots,m_n] U[m_1]\cdots U[m_n] e^{j2\pi(m_1+\cdots+m_n)k/N} \\
&= \frac{1}{N} \sum_{m=0}^{N-1} \left\{ \frac{1}{N^{n-1}} \sum_{m_1=0}^{N-1} \cdots \sum_{m_{n-1}=0}^{N-1} Y_n[m_1,\cdots,m_{n-1},m-m_1-\cdots-m_{n-1}] \right\} e^{j2\pi mk/N} \\
&= \frac{1}{N} \sum_{m=0}^{N-1} Y_n[m] e^{j2\pi mk/N}
\end{aligned} \quad (44)$$

where

$$Y_n[m] = \frac{1}{N^{n-1}} \sum_{m_1=0}^{N-1} \cdots \sum_{m_{n-1}=0}^{N-1} Y_n[m_1,\cdots,m_{n-1},m-m_1-\cdots-m_{n-1}] \quad (45)$$

$$Y_n[m_1,\cdots,m_n] = H_n[m_1,\cdots,m_n] U[m_1]\cdots U[m_n] \quad (46)$$

Let

$$U_n[m] = \frac{1}{N^{n-1}} \sum_{m_1=0}^{N-1} \cdots \sum_{m_{n-1}=0}^{N-1} U[m_1]\cdots U[m_{n-1}] U[m-m_1-\cdots-m_{n-1}] \quad (47)$$

The $n$th-order discrete quasi-linear transfer function (DQLTF) can then be defined as

$$G_n[m] = \frac{Y_n[m]}{U_n[m]}, \quad m \in M_n \quad (48)$$

where $M_n$ is valid domain for $n$th-order discrete output frequency function $Y_n[m]$.

## 6. Determination of the output frequency range

It is observed that the output frequency components of nonlinear systems are much richer compared to the corresponding input frequencies. The input frequencies will pass in a coupled way through a nonlinear system, that is, an input at given frequencies may produce quite different output frequencies. Therefore energy may be transferred to or from other frequency components. This is quite different from the case for linear systems where the output frequency range is identical in steady state to that of the inputs. It would be difficult to give a general explicit expression connecting the input and output frequencies for most nonlinear systems. However, for some special cases, where the input and output frequencies of the nonlinear system are restrained by some conditions, explicit algorithms are available to determine the output frequency range (Lang and Billings 1996, 1997). In this section, new algorithms, which can be used for determining the output frequency range of arbitrary order quasi-linear transfer function $G_n(j\omega)$ described in Sections 3 and 4, are proposed by adapting and improving the algorithms presented by Lang and Billings (1996, 1997, 2000).

6.1 The frequency range of the quasi-linear transfer functions for a general input

From the viewpoint of practical applications, a realistic assumption for the input spectrum is

$$U(j\omega) = \begin{cases} U_0(j\omega) & a \leq |\omega| \leq b \\ 0 & \text{elsewhere} \end{cases} \tag{49}$$

where $0 \leq a \leq b \leq \infty$ represents the real frequency range of the input signal. Under this assumption, it is clear by observing the variable transform (11) that the input and output frequencies satisfy

$$\omega = \sum_{i=1}^{n} \omega_i, \quad \omega_i \in [-b,-a] \cup [a,b] \tag{50}$$

For the case of the first-order nonlinear subsystem where $n=1$, it is clear that the output frequency range is $\omega \in \Omega_1 = [-b,-a] \cup [a,b]$. For the case of $n=2$, noticing that the roles of $\omega_1$ and $\omega_2$ in (50) are symmetrical, the output frequency range can be determined by inspecting the three combinations listed in Table 1, where the set $\Omega_2$ indicates the frequency range.

Table 1 Output frequency range for the case of $n=2$

| $\omega_1$ | $\omega_2$ | $\omega$ |
|---|---|---|
| [a, b] | [a, b] | [2a, 2b] |
| [a, b] | [-b, -a] | [a-b, b-a] |
| [-b, -a] | [-b, -a] | [-2b, -2a] |
| $\Omega_2 = [-2b,-2a] \cup [a-b,b-a] \cup [2a,2b]$ |||

Table 2 Output frequency range for the case of $n=3$

| $\omega_1$ | $\omega_2$ | $\omega_3$ | $\omega$ |
|---|---|---|---|
| [a, b] | [a, b] | [a, b] | [3a, 3b] |
| [a, b] | [a, b] | [-b, -a] | [2a-b, 2b-a] |
| [a, b] | [-b, -a] | [-b, -a] | [a-2b, b-2a] |
| [-b, -a] | [-b, -a] | [-b, -a] | [-3b, -3a] |
| $\Omega_3 = [-3b,-3a] \cup [a-2b,b-2a] \cup [2a-b,2b-a] \cup [3a,3b]$ ||||

Similarly, for the case of $n=3$, the output frequency range can be determined by inspecting four combinations and these are listed in Table 2, where the set $\Omega_3$ indicates the frequency range for this case.

Let $\Omega_n$ denote the frequency range of the $n$th-order quasi-linear transfer function $G_n(j\omega)$ described in Sections 3 and 4. The algorithm for calculating $\Omega_n$ can be summarised below:

**Proposition 6.1** *Assume that a nonlinear system is excited by a signal with a band-limited spectrum u(t), whose spectrum is defined as in (44). The frequency range $\Omega_n$ of the nth-order quasi-linear transfer function $G_n(j\omega)$ is given by*

$$\Omega_n = \bigcup_{k=0}^{n}[a_k, b_k], \quad n=1,2,\cdots \tag{51}$$

*where*

$$a_k = ka - (n-k)b \tag{52}$$

$$b_k = kb - (n-k)a, \quad k=0,2,\cdots,n \tag{53}$$

To illustrate the application of Proposition 6.1, consider a simple example where $a=20$, $b=50$. The sets $\Omega_n$ for $n=1,2,3$ can be calculated to be:

For $n=1$, $\Omega_1 = [-50,-20] \cup [20,50]$.

For $n=2$, $a_0 = -100$, $b_0 = -40$, $a_1 = -30$, $b_1 = 30$, $a_2 = 40$, $b_2 = 100$. Therefore $\Omega_2 = [-100,-40] \cup [-30,30] \cup [40,100]$.

For $n=3$, $a_0 = -150$, $b_0 = -60$, $a_1 = -80$, $b_1 = 10$, $a_2 = -10$, $b_2 = 80$, $a_3 = 60$, $b_3 = 150$. Therefore, $\Omega_3 = [-150,-60] \cup [-80,10] \cup [-10,80] \cup [60,150] = [-150,150]$.

It can be seen that the frequency range $\Omega_n$ of the $n$th-order QLTF is always symmetric about the original. The non-negative range $\Omega_n^+$ is given below:

**Proposition 6.2** *Assume that a nonlinear system is excited by signal with a band-limited spectrum u(t), whose spectrum is defined as in (49). The non-negative frequency range $\Omega_n^+$ of the nth-order quasi-linear transfer function $G_n(j\omega)$ is given by*

$$\Omega_n^+ = \bigcup_{k=0}^{n}[a_k', b_k'], \quad n=1,2,\cdots \tag{54}$$

*where*

$$a_k' = \max\{a_k, 0\}, \tag{55}$$

$$b_k' = \max\{b_k, 0\}, \tag{56}$$

$$a_k = ka - (n-k)b, \tag{57}$$

$$b_k = kb - (n-k)a, \quad n \geq 1 \text{ and } k = 0,1,\cdots,n-1. \tag{58}$$

Set $a=20$, $b=50$. For the case $n=2$, it can be calculated that $a'_0 = b'_0 = 0$, $a'_1 = 0$, $b'_1 = 30$, $a'_2 = 40$, $b'_2 = 100$. Therefore, $\Omega_2^+ = \{0\} \cup [0,30] \cup [40,100] = [0,30] \cup [40,100]$.

6.2 The frequency range of the quasi-linear transfer functions for a multitone input

This has been discussed in Lang and Billings (1996), where an algorithm has been proposed to compute the frequency range of arbitrary order input/output frequency function $Y_n(j\omega)$ defined by (14). In this study, a much more smarter and compact recursive algorithm is proposed for calculating the frequency range of arbitrary order QLTF.

From Eqs. (30) and (32), the input and output frequencies for the $n$th-order subsystem with a multitone input are constrained by

$$\omega = \sum_{i=1}^{n} \omega_{k_i}, \ k_i \in \{\pm 1, \pm 2, \cdots, \pm K\} \tag{59}$$

This will be used to determine the frequency range of the $n$th-order QLTF. Observing that the roles of $\omega_{k_i}$ for $i=1,2,\ldots,n$ are symmetric, only the non-negative frequency components of the output frequency $\omega$ need to be determined.

For the simplest case of $n=1$, it is clear that the frequency range of the output spectrum is $\omega \in \Omega_1^+ = \{\omega_k : k = 1, \cdots, K\}$. In order to determine the non-negative frequency range $\Omega_2^+$ for the case of $n=2$, inspect first the following combinations of two frequency components

$$\begin{Bmatrix} \omega_1 - \omega_K \\ \vdots \\ \omega_1 - \omega_1 \\ \omega_1 + \omega_1 \\ \vdots \\ \omega_1 + \omega_K \\ \vdots \\ \omega_K - \omega_K \\ \vdots \\ \omega_K - \omega_1 \\ \omega_K + \omega_1 \\ \vdots \\ \omega_K + \omega_K \end{Bmatrix} \tag{60}$$

This can be expressed in a vector form as

$$\Gamma_2 = \begin{bmatrix} \omega_1 I_{2K} \\ \vdots \\ \omega_K I_{2K} \end{bmatrix} + \begin{bmatrix} V \\ \vdots \\ V \end{bmatrix} = \begin{bmatrix} \omega_1 \\ \vdots \\ \omega_K \end{bmatrix} \otimes \begin{bmatrix} I_{2K} \\ \vdots \\ I_{2K} \end{bmatrix} + \begin{bmatrix} V \\ \vdots \\ V \end{bmatrix} \tag{61}$$

where $I_{2K} = \underbrace{[1,\cdots,1]}_{2K}{}^T$, $V = [-\omega_K,\cdots,-\omega_1,\omega_1,\cdots,\omega_K]^T$. The symbol '$\otimes$' denotes the Kronecker product, which is defined as

$$A_{p\times 1} \otimes B_{q\times 1} = \begin{bmatrix} a_1 B \\ a_2 B \\ \cdots \\ a_p B \end{bmatrix} \tag{62}$$

For a given vector $X = [x_1, x_2, \cdots, x_p]^T$, let $<X>$ denote a set whose elements are formed by the entities of $X$ in the sense that $<X> = \{|x_i| : 1 \leq i \leq p\}$. It can be easily proved that all the different entities of the vector $<\Gamma_2>$ are identical to all the non-negative frequency components of the second order QLTF. Note that some entities in the vector $<\Gamma_2>$ may be the same. Therefore $<\Gamma_2>$ is redundant for computing the non-negative frequency components of the second order QLTF.

In general, the non-negative frequency components of the $n$th order QLTF can be calculated using the recursive algorithm below:

**Proposition 6.3** *Assume that a nonlinear system is excited by multitone signal u(t) with K fundamental frequency components, $W = [\omega_1, \omega_2, \cdots, \omega_K]^T$. The non-negative frequency components of the nth order QLTF can be determined by searching all the different entities of $\Omega_n^+$, which is defined as*

$$\Gamma_1 = W \tag{63}$$

$$\Gamma_n = \Gamma_{n-1} \otimes I_{2K} + I_{(2K)^{n-2}} \otimes G \tag{64}$$

$$\Omega_n^+ = <\Gamma_n>, \ n \geq 2 \tag{65}$$

*where*

$$I_m = \underbrace{[1,1,\cdots,1]}_{m}{}^T \tag{66}$$

$$G = I_K \otimes V = \underbrace{[V,V,\cdots,V]}_{K}{}^T \tag{67}$$

$$V = [-\omega_K,\cdots,-\omega_1,\omega_1,\cdots,\omega_K]^T \tag{68}$$

The above recursive algorithm is very simple and quite easy to realise using vector-oriented software tools. As an example, consider the case of $K=3$, $W = [2,3,7]^T$. For $n=2$ and 3, the non-negative frequency components of the QLTFs were calculated to be $\Omega_2^+ = \{0,1,4,5,6,9,10,14\}$ and $\Omega_3^+ = \{1,2,3,4,6,7,8,9, 11,12,13,16,17,21\}$.

## 7. Simulation studies

An example relating to a continuous-time nonlinear system will be studied to illustrate the physical meaning of the proposed quasi-linear transfer functions. Only the case of a multitone input will be considered in this example.

Consider a nonlinear system described by the equation

$$M\frac{d^2 y(t)}{dt^2} + D\frac{dy(t)}{dt} + K_1 y(t) + K_2 y^2(t) + K_3 y^3(t) = u_0(t) \tag{69}$$

where $u_0(t)$ is the force imposed on a mass $M$, $y(t)$ is the displacement of the mass relative to the equilibrium position, $D$ is the damping constant, $K_1, K_2$ and $K_3$ are the constants of stiffness relative to the displacement and the square and cubic of the displacement, respectively. By setting

$$\zeta = \frac{D}{2\sqrt{MK_1}}, \quad \omega_n = \sqrt{\frac{K_1}{M}}, \quad \varepsilon_2 = \frac{K_2}{K_1}, \quad \varepsilon_3 = \frac{K_3}{K_1}$$

Equation (69) can be expressed in a standard form

$$\frac{d^2 y(t)}{dt^2} + 2\zeta\omega_n \frac{dy(t)}{dt} + \omega_n^2 y(t) + \varepsilon_2 \omega_n^2 y^2(t) + \varepsilon_3 \omega_n^2 y^3(t) = u(t) \tag{70}$$

In this example, it was assumed that $\omega_n = 10$, $\zeta = 0.1$, $\varepsilon_2 = 10^3$, $\varepsilon_3 = 5\times 10^5$. This model includes both a square and a cubic nonlinearity. The first order frequency response function can easily be determined from the linear part of the equation as

$$H_1(j\omega) = \frac{1}{(j\omega)^2 + 2\zeta\omega_n(j\omega) + \omega_n^2} \tag{71}$$

The second and third order frequency response functions for this system were derived (Biilings and Peyton Jones 1990) as

$$H_2(j\omega_1, j\omega_2) = -\varepsilon_2 \omega_n^2 H_1(j\omega_1) H_1(j\omega_2) H_1(j\omega_1 + j\omega_2) \tag{72}$$

$$H_3(j\omega_1, j\omega_2, j\omega_3) = -\frac{1}{6}\omega_n^2 \{4\varepsilon_2 [H_1(j\omega_1) H_2(j\omega_2, j\omega_3) + H_1(j\omega_2) H_2(j\omega_3, j\omega_1)$$
$$+ H_1(j\omega_3) H_2(j\omega_1, j\omega_2)] + 6\varepsilon_3 H_1(j\omega_1) H_1(j\omega_2) H_1(j\omega_3)\}$$
$$\times H_1(j\omega_1 + j\omega_2 + j\omega_3) \tag{73}$$

The meaning of the generalised frequency response functions can be interpreted in a graphical way (Peyton Jones and Billings 1990, Zhang and Billings 1993). In this example, however, the quasi-linear transfer functions will be used to interpret the frequency characteristics of the nonlinear system. Consider the two-tone input

$$u(t) = 0.25\cos(2.5t) + 0.75\cos(7.5t) \tag{74}$$

where $A_0 = 0$, $A_{-1} = A_1 = 0.25$, $A_{-2} = A_2 = 0.75$, $\omega_0 = 0$, $\omega_{-1} = -\omega_1 = -2.5$, $\omega_{-2} = -\omega_2 = -7.5$.

The frequency response function of the linear part of the system is given by (71), for which the output frequencies are the same as the input frequencies, that is, $\Omega_1 = \{2.5, 7.5\}$. Using Proposition 6.3, the non-negative output frequency components for the output of the second and third order QLTFs were calculated to be $\Omega_2^+ = \{0, 5, 10, 15\}$ and $\Omega_3^+ = \{2.5, 7.5, 12.5, 17.5, 22.5\}$. Thus, the whole components of the output frequencies of the second and third order QLTFs should therefore be, respectively

$$\Omega_2 = -\Omega_2^+ \cup \Omega_2^+ = \{-15, -10, -5, 0, 5, 10, 15\} \tag{75}$$

$$\Omega_3 = -\Omega_3^+ \cup \Omega_3^+ = \{-22.5, -17.5, -12.5, -7.5, -2.5, 2.5, 7.5, 12.5, 17.5, 22.5\} \tag{76}$$

From Eqs. (30) and (32)

$$\frac{2}{\pi} U_2(j\omega) = \sum_{k_1=-2}^{2} B(\omega_{k_1}) B(\omega - \omega_{k_1})$$

$$= B(\omega_{-2}) B(\omega - \omega_{-2}) + B(\omega_{-1}) B(\omega - \omega_{-1}) + B(\omega_1) B(\omega - \omega_1) + B(\omega_2) B(\omega - \omega_2)$$

$$= B(-\omega_2) B(\omega + \omega_2) + B(-\omega_1) B(\omega + \omega_1) + B(\omega_1) B(\omega - \omega_1) + B(\omega_2) B(\omega - \omega_2) \tag{77}$$

$$\frac{2}{\pi} Y_2(j\omega) = \sum_{k_1=-2}^{2} B(\omega_{k_1}) B(\omega - \omega_{k_1}) H_2(j\omega_{k_1}, j(\omega - \omega_{k_1}))$$

$$= B(-\omega_2) B(\omega + \omega_2) H_2(-j\omega_2, j(\omega + \omega_2)) + B(-\omega_1) B(\omega + \omega_1) H_2(-j\omega_1, j(\omega + \omega_1))$$

$$+ B(\omega_1) B(\omega - \omega_1) H_2(j\omega_1, j(\omega - \omega_1)) + B(\omega_2) B(\omega - \omega_2) H_2(j\omega_2, j(\omega - \omega_2)) \tag{78}$$

Noting the definition of (31), $U_2(j\omega)$ and $Y_2(j\omega)$ for $\omega \in \Omega_2$ can be calculated as

$U_2(j(-15)) = 0.5\pi B(-7.5) B(-7.5)$

$Y_2(j(-15)) = 0.5\pi B(-7.5) B(-7.5) H_2(j(-7.5), j(-7.5))$

$U_2(j(-10)) = 0.5\pi [B(-7.5) B(-2.5) + B(-2.5) B(-7.5)]$

$Y_2(j(-10)) = 0.5\pi [B(-7.5) B(-2.5) H_2(j(-7.5), j(-2.5)) + B(-2.5) B(-7.5) H_2(j(-2.5), j(-7.5))]$

$U_2(j(-5)) = 0.5\pi [B(-7.5) B(2.5) + B(-2.5) B(-2.5) + B(2.5) B(-7.5)]$

$Y_2(j(-5)) = 0.5\pi [B(-7.5) B(2.5) H_2(j(-7.5), j2.5)$
$\qquad + B(-2.5) B(-2.5) H_2(j(-2.5), j(-2.5)) + B(2.5) B(-7.5) H_2(j2.5, j(-7.5))]$

$U_2(j0) = 0.5\pi [B(-7.5) B(7.5) + B(-2.5) B(2.5) + B(2.5) B(-2.5) + B(7.5) B(-7.5)]$

$Y_2(j0) = 0.5\pi [B(-7.5) B(7.5) H_2(j(-7.5), j7.5) + B(-2.5) B(2.5) H_2(j(-2.5), j2.5)]$
$\qquad + B(2.5) B(-2.5) H_2(j2.5, j(-2.5)) + B(7.5) B(-7.5) H_2(j7.5, j(-7.5))]$

$U_2(j5) = 0.5\pi [B(-2.5) B(7.5) + B(2.5) B(2.5) + B(7.5) B(-2.5)]$

$Y_2(j5) = 0.5\pi [B(-2.5) B(7.5) H_2(j(-2.5), j7.5)$
$\qquad + B(2.5) B(2.5) H_2(j2.5, j2.5) + B(-2.5) B(7.5) H_2(j(-2.5), j7.5)]$

$U_2(j10) = 0.5\pi [B(2.5) B(7.5) + B(7.5) B(2.5)]$

$$Y_2(j10) = 0.5\pi[B(2.5)B(7.5)H_2(j2.5, j7.5) + B(7.5)B(2.5)H_2(j7.5, j2.5)]$$

$$U_2(j15) = 0.5\pi B(7.5)B(7.5)$$

$$Y_2(j15) = 0.5\pi B(7.5)B(7.5)H_2(j7.5, j7.5)$$

Finally, from the definition (19), the magnitudes and phase angles of the second-order QLTF at different frequency components were calculated and are listed in Table 3. These were also plotted in Figure 1. The third-order QLTF were also calculated in the same way and is shown in Figure 2.

The QLTFs of the system (70) with the two-tone input (74) reveals the following facts:

i) The magnitude of the QLTFs are symmetrical about the origin.

ii) The phase angle of the QLTFs are anti-symmetrical about the origin.

iii) The energy at the input frequencies $\omega_1 = 2.5$ and $\omega_1 = 7.5$ was transferred to the output at different frequencies.

iv) The magnitude of the second QLTF reaches a maximum at the frequencies $\omega = \pm 10$, which are modulated from the input frequencies. This is reasonable by observing that natural frequency of the system was $\omega_n = 10$. The QLTFs therefore clearly detect the resonant peak of the system spectrum.

v) It has been noted that in some cases the system output in the time domain cannot sensitively reflect the changes of a particular parameter relating to an identified time-domain model. The QLTFs, however, can sensitively reflect the changes of some structure parameters that cannot be detected by time-domain model.

To illustrate the final point above, consider the model (70). By setting the input as in (74), this model was simulated in the interval [0, 10] using a Runge-Kutta algorithm with an integral step $h = 0.005$ and initial condition $y(0) = \dot{y}(0) = 0$. To inspect how the parameter $\varepsilon_2$ affects the system output behaviour in the time domain, two groups of structure parameters were considered in the simulation:

(I) $\omega_n = 10$, $\zeta = 0.1$, $\varepsilon_2 = 10^3$, $\varepsilon_3 = 5 \times 10^5$.

(II) $\omega_n = 10$, $\zeta = 0.1$, $\varepsilon_2 = 10^2$, $\varepsilon_3 = 5 \times 10^5$.

To inspect the effect of the parameter $\varepsilon_2$ on the system output behaviour in the time domain, the system outputs about the position $y(t)$ and the velocity $\dot{y}(t)$ with the two groups of parameters were compared and are shown in Figures 3 and 4. The 2D-projections of attractors and the phase portraits were also considered and are shown in Figures 5 and 6. By inspecting these figures, it can be concluded that the effect of the change of the parameter $\varepsilon_2$ on the system output in the time domain is not very clear or obvious. As will be seen, the changes of the system structure parameter, however, can be sensitively detected by the proposed QLTFs.

Table 3 The second order quasi-linear transfer function of the system (70) with the two-tone input (74)

| $\omega_k$ | $|G_2(j\omega_k)|$ | $\angle G_2(j\omega_k)$ |
|---|---|---|
| -15 | 0.3637 | 24.35° |
| -10 | 1.1515 | -68.02° |
| -5 | 0.2820 | -157.27° |
| 0 | 0.4321 | 180° |
| 5 | 0.2820 | 157.27° |
| 10 | 1.1515 | 68.02° |
| 15 | 0.3637 | -24.35° |

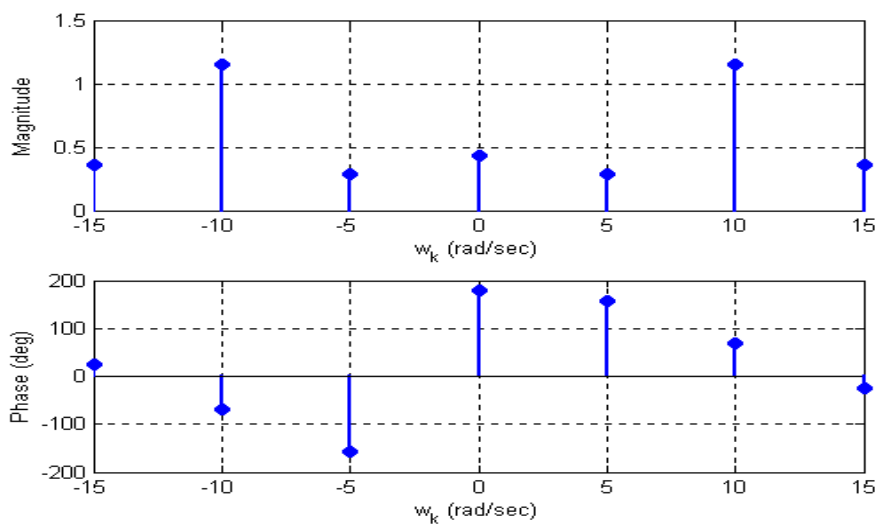

Figure 1  The magnitude and phase of the second order quasi-linear transfer function $G_2(j\omega)$ of the system (70) with the two-tone input (74) and the structure parameters $\omega_n = 10$, $\zeta = 0.1$, $\varepsilon_2 = 10^3$, $\varepsilon_3 = 5 \times 10^5$.

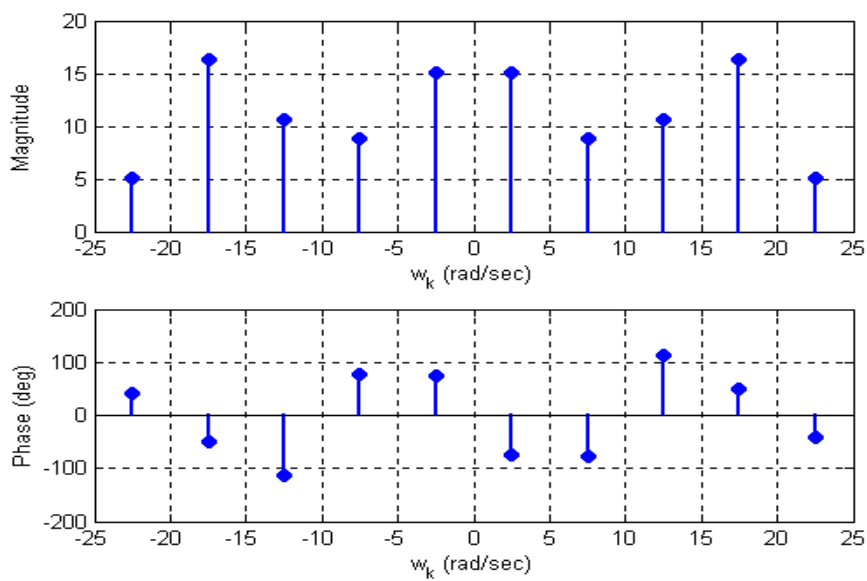

Figure 2  The magnitude and phase of the third order quasi-linear transfer function $G_3(j\omega)$ of the system (70) with the two-tone input (74) and the structure parameters $\omega_n = 10$, $\zeta = 0.1$, $\varepsilon_2 = 10^3$, $\varepsilon_3 = 5 \times 10^5$.

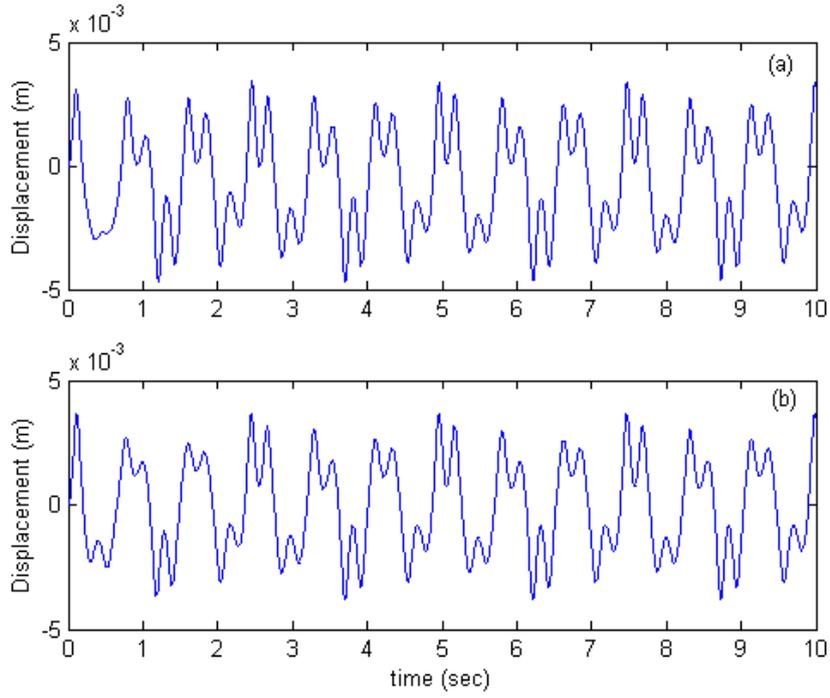

Figure 3  The system output (displacement) of the equation (70) with the input (74).
(a) $\omega_n = 10$, $\zeta = 0.1$, $\varepsilon_2 = 10^3$, $\varepsilon_3 = 5 \times 10^5$. (b) $\omega_n = 10$, $\zeta = 0.1$, $\varepsilon_2 = 10^2$, $\varepsilon_3 = 5 \times 10^5$.

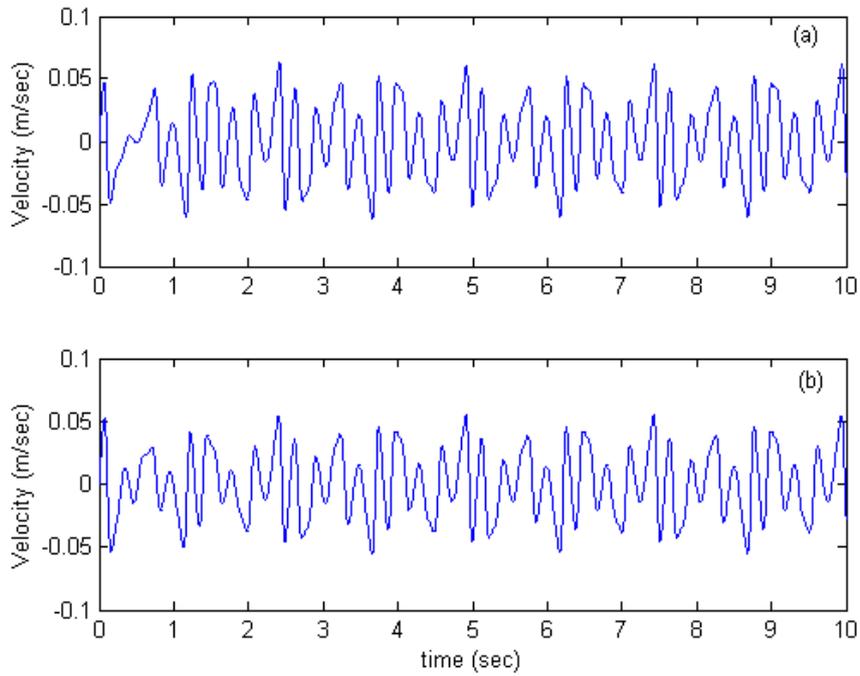

Figure 4  The system output (velocity) of the equation (70) with the input (74).
(a) $\omega_n = 10$, $\zeta = 0.1$, $\varepsilon_2 = 10^3$, $\varepsilon_3 = 5 \times 10^5$. (b) $\omega_n = 10$, $\zeta = 0.1$, $\varepsilon_2 = 10^2$, $\varepsilon_3 = 5 \times 10^5$.

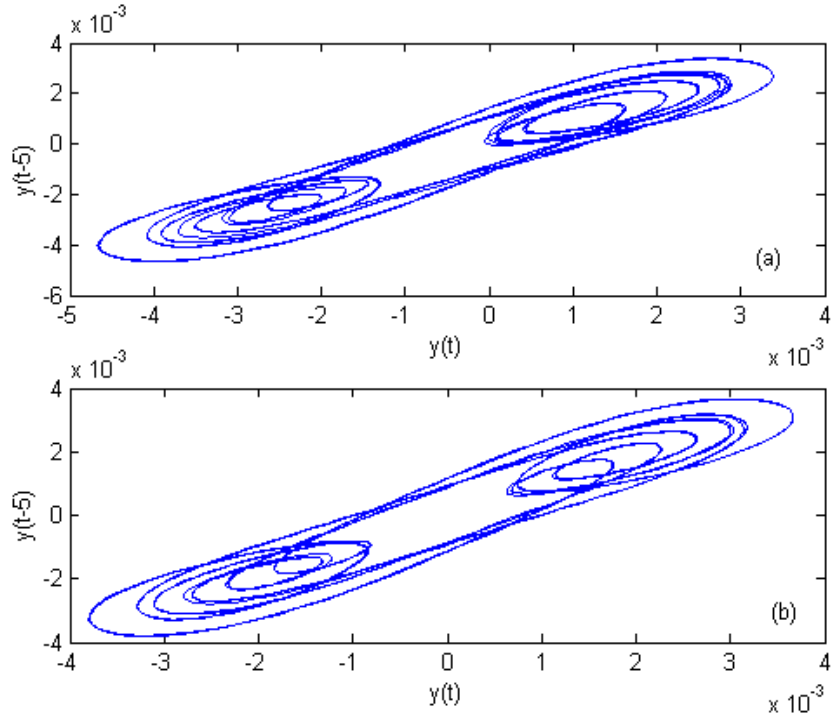

Figure 5  Phase portraits for the equation (70) with the input (74).
(a) $\omega_n = 10$, $\zeta = 0.1$, $\varepsilon_2 = 10^3$, $\varepsilon_3 = 5 \times 10^5$. (b) $\omega_n = 10$, $\zeta = 0.1$, $\varepsilon_2 = 10^2$, $\varepsilon_3 = 5 \times 10^5$.

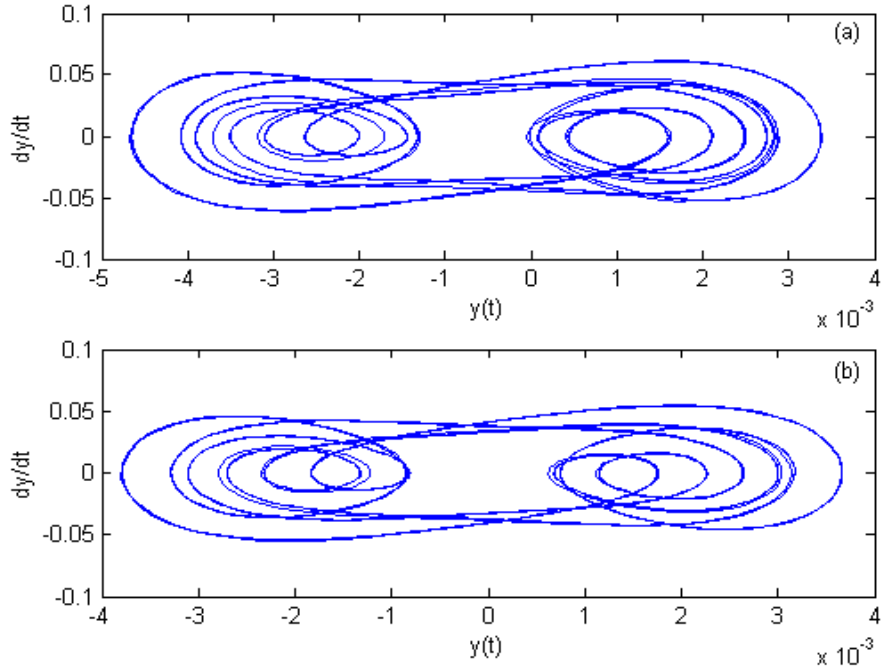

Figure 6  2D-projections of attractors for the equation (70) with the input (74).
(a) $\omega_n = 10$, $\zeta = 0.1$, $\varepsilon_2 = 10^3$, $\varepsilon_3 = 5 \times 10^5$. (b) $\omega_n = 10$, $\zeta = 0.1$, $\varepsilon_2 = 10^2$, $\varepsilon_3 = 5 \times 10^5$.

Now, consider again the second and third order QLTFs by setting $\varepsilon_2 = 10^2$. The resulting second order QLTF is shown in Figure 7, where it can be seen that the magnitude of the transfer function at all the frequency components have been dramatically compressed, whereas the phase angles remain the same. The resulting third order QLTF is shown in Figure 8, where both the magnitude and phase angle are greatly different from those shown in Figure 2. The sensitivity of the third order QLTF to the structure parameter $\varepsilon_3$ can also be seen by comparing Figure 9 with Figure 2, where the parameter $\varepsilon_3$ was set be $2.5 \times 10^5$ and $5 \times 10^5$, respectively. It is clear by comparing these figures that the QLTFs can be very sensitive to changes of the model parameters. This should be so, since the parameters $\varepsilon_2$ and $\varepsilon_3$ directly affect the frequency response functions as shown in (72) and (73).

It can be seen from this example that QLTFs not only characterise the output frequency behaviour but may also sensitively detect changes of model parameters of nonlinear systems. As a consequence, this example suggests that the proposed concept of QLTFs may be applied in fault detection and diagnosis for engineering systems. For example, the ideal frequency characteristics of the underlying system can be documented and stored in a database. The frequency behaviour of the system being operated in the real environment can be analysed using the concept of QLTFs. A decision can then be made by comparing the frequency behaviour of the real system with the known ideal characteristics. In this way, faults that arise in the real system should be detected.

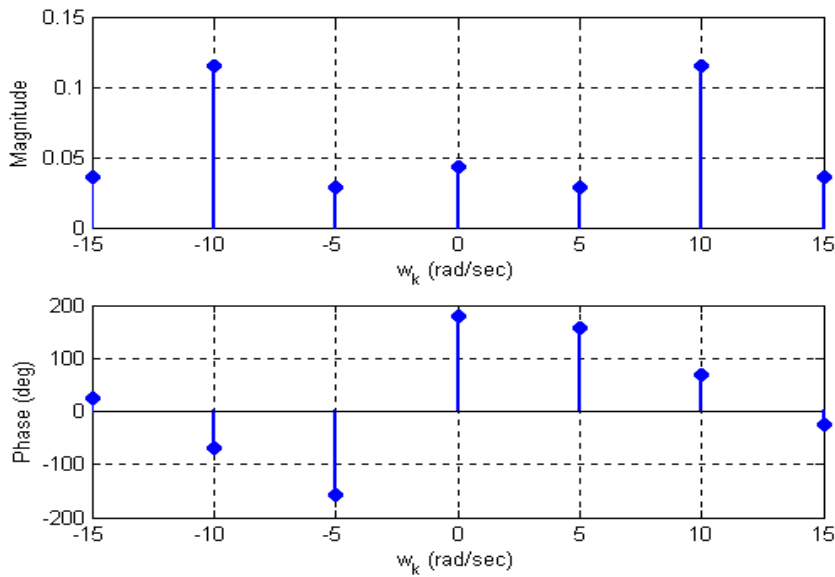

Figure 7  The magnitude and phase of the second order quasi-linear transfer function $G_2(j\omega)$ of the system (70) with the two-tone input (74) and the structure parameters $\omega_n = 10$, $\zeta = 0.1$, $\varepsilon_2 = 10^2$, $\varepsilon_3 = 5 \times 10^5$.

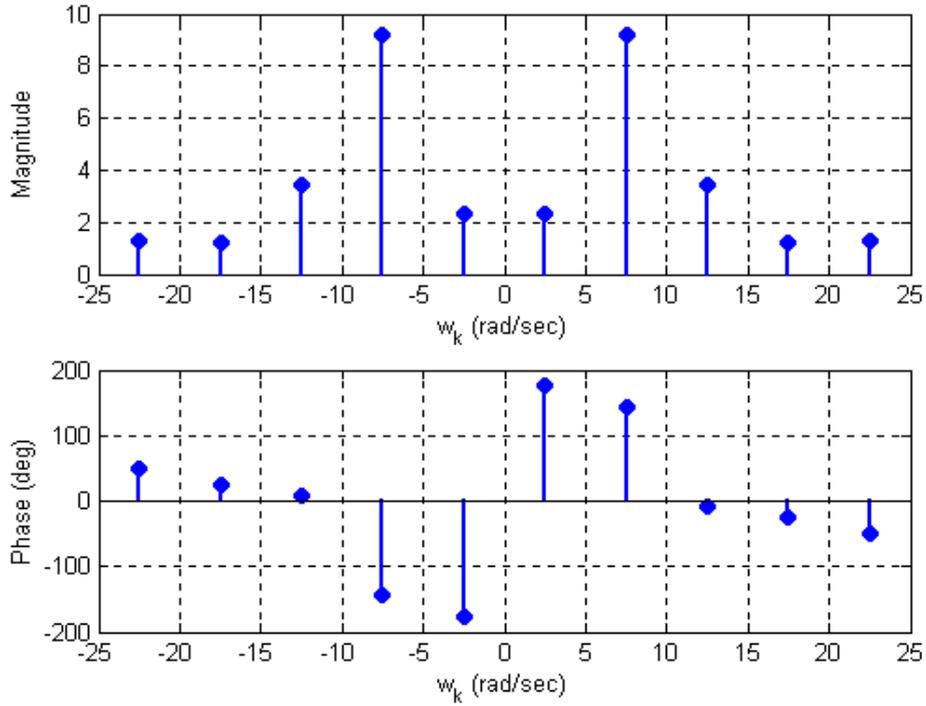

Figure 8  The magnitude and phase of the third order quasi-linear transfer function $G_3(j\omega)$ of the system (70) with the two-tone input (74) and the structure parameters $\omega_n = 10$, $\zeta = 0.1$, $\varepsilon_2 = 10^2$, $\varepsilon_3 = 5 \times 10^5$.

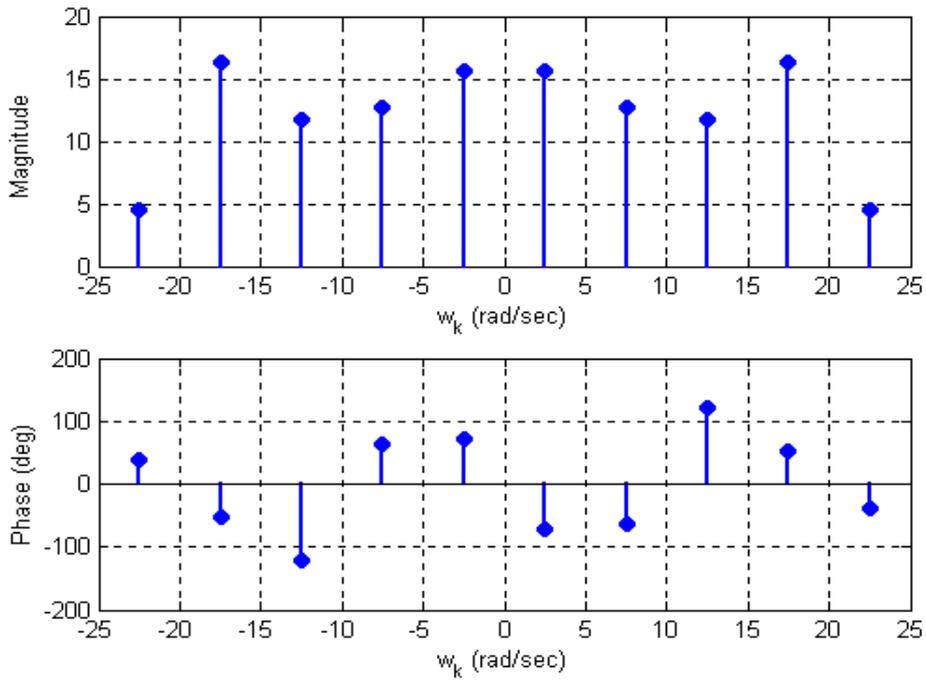

Figure 9  The magnitude and phase of the third order quasi-linear transfer function $G_3(j\omega)$ of the system (70) with the two-tone input (74) and the structure parameters $\omega_n = 10$, $\zeta = 0.1$, $\varepsilon_2 = 10^3$, $\varepsilon_3 = 2.5 \times 10^5$.

## 8. Conclusions

The concept of quasi-linear transfer functions (QLTFs) has been introduced to characterise the output frequency behaviour of nonlinear systems. The newly defined QLTFs were derived from the well known Volterra series representation and the generalised frequency response functions (GFRFs). QLTFs are one-dimensional functions, which are similar in form to the traditional transfer functions for linear systems. Although OLTFs are of a simple form compared with conventional GFRFs, they contain most of the information that is available in GFRFs but represent this in a different way. QLTFs can therefore be used to characterise the output frequency behaviour of nonlinear systems that can be described by a Volterra series. It has been shown that QLTFs not only characterise the output frequency behaviour but may also detect changes in the model parameters. A potential application of QLTFs is therefore in structure detection and fault diagnosis for nonlinear engineering systems.

The emphasis of this study has been to introduce the concept of OLTFs. This new concept was also extended from the continuous-time description to the case of discrete-time systems. An explicit expression between the input and output frequency responses has also been derived to reveal how the output frequencies of a nonlinear system rely on the input frequencies. New algorithms for estimating the frequency range of arbitrary order nonlinear outputs have also been proposed by considering the two cases of a multitone and a general input excitation.

In the example, the QLTFs were computed based on the estimated GFRFs. It would however be desirable to estimate the QLTFs directly from input-output observations. This will be one of the topics of the future studies.

## Acknowledgements
The authors gratefully acknowledge that this work was supported by EPSRC(UK).

## References

Barrett, J., 1963, The use of functionals in the analysis of nonlinear physical systems. *Journal of Electronics and Control*, **15**, 567-615.

Bedrosian, E. and Rice, S.O., 1971, The output properties of Volterra systems driven by harmonic and Gaussian inputs. *Proceedings of the Institute of Electrical and Electronics Engineers* (*IEEE*), **59**, 1688-1707.

Billings, S.A., 1980, Identification of nonlinear systems—a survey. *Proceedings of IEEE*, **127**, 272-285.

Billings, S.A., and Lang, Z.Q., 1996, A bound for the magnitude characteristics of nonlinear output frequency response functions—Part 1: Analysis and computation. *International Journal of Control*, **65**, 309-328.

Billings, S.A., and Peyton Jones, J.C., 1990, Mapping nonlinear intero-differential equation into the frequency domain. *International Journal of Control*, **52**, 863-879.

Billings, S.A., and Tsang, K.M., 1989a, Spectral analysis for nonlinear systems—Part I: Parametric nonlinear spectral analysis. *Mechanical Systems and Signal Processing*, **3**, 319-339; 1989b, Spectral analysis for nonlinear systems—Part II: Interpretation of nonlinear frequency response functions. *Mechanical Systems and Signal Processing*, **3**, 341-359; 1990, Spectral analysis for nonlinear systems—Part III: Case study examples. *Mechanical Systems and Signal Processing*, **4**, 3-21.

Billings, S.A., and Yusof, M.I., 1996, Decomposition of generalised frequency response functions for nonlinear systems using symbolic computation. *International Journal of Control*, **65**, 589-618.

Boaghe, O.M., Balikhin, M.A., Billings, S.A., and Alleyne, H.(2001), Identification of nonlinear processes in the magnetosphere dynamics and forecasting of *Dst* index, *Journal of Geophysical Research-Space Physics*, **106(A12)**, 30047-30066.

Boyd, S.P. and Chua, L.O.(1985). Fading memory and the problem of approximating nonlinear operators with Volterra series. *IEEE Transactions on Circuits and Systems*(**CAS**), **32**, 1150-1161.

Brilliant, M., 1958, Theory of the analysis of nonlinear systems. *Technical Report No. 345*, *Research Laboratory of Electronics* (*RLE*), *Massachusetts Institute of Technology* (*MIT*).

Bussgang, J.J., Ehrman, L., and Garham, J.W., 1974, Analysis of nonlinear systems with multiple inputs. *Proceedings of IEEE*, **62**, 1088-1119.

Chua, L.O., and Ng, C.Y., 1979a, Frequency domain analysis of nonlinear systems: general theory. *IEE Journal Electronic Circuits and Systems*, **3**, 165-185; 1979b, Frequency domain analysis of nonlinear systems: formulation of transfer functions. *IEE Journal Electronic Circuits and Systems*, **3**, 257-269.

George, D., 1959, Continuous nonlinear systems. *Technical Report No. 355*, *Research Laboratory of Electronics* (*RLE*), *Massachusetts Institute of Technology* (*MIT*).

Gifford, S.J., and Tomlinson, G.R., 1989, Recent advances in the application of functional series to nonlinear structure. *Journal of Sound and Vibration*, **135**, 289-317.

Lang, Z.Q., and Billings, S.A., 1996, Output frequency characteristics of nonlinear systems. *International Journal of Control*, **64**, 1049-1067.

Lang, Z.Q., and Billings, S.A., 1997, Output frequencies of nonlinear systems. *International Journal of Control*, **67**, 713-730.

Lang, Z.Q., and Billings, S.A., 2000, Evaluation of output frequency response of nonlinear systems under multiple inputs. *IEEE Transactions on Circuits and Systems*, **3**, 165-185.

Leontaritis,I.J., Billings,S.A., 1985, Input-output parametric models foe non-linear systems, part I: deterministic non-linear systems; part II: stochastic non-linear systems. *International Journal of Control*, **41(2)**,303-344.

Marmarelis, P., and Marmarelis, V., 1978, *Analysis of Physiological Systems*. New York: Plenum.

Masri, S.F., and Caughey, T.K., 1979, A non-parametric identification technique for non-linear dynamic problems. *Journal of Applied Mechanics*, **46**, 433-447.

Palumbo, P., and Piroddi, L., 2000, Harmonic analysis of non-linear structures by means of generalised frequency response functions coupled wirh NARX models. *Mechanical Systems and Signal Processing*, **14**, 243-265.

Peyton Jones, J.C., and Billings, S.A., 1989, Recursive algorithm for computing the frequency response of a class of nonlinear difference equation models. *International Journal of Control*, **50**, 1925-1940.

Peyton Jones, J.C., and Billings, S.A., 1990, Interpretation of nonlinear frequency response functions. *International Journal of Control*, **52**, 319-346.

Powers, E.J., and Miksad, R.W., 1987, Polyspectral measurement and analysis of nonlinear wave interactions. In *Proceedings of the Symposium on Nonlinear Wave Interactions in Fluids*, Boston, USA, pp. 9-16.

Rugh, W.J., 1981, *Nonlinear System Theory—the Voterra/Wiener Approach*. The John Hopkins University Press.

Schetzen, M., 1980a, *The Volterra and Wiener Theories of Nonlinear Systems*. New York: John Wiley & Sons.

Schetzen, M., 1980b, Measurement of the kernels of a system of finite order. *International Journal of Control*, **1**, 251-263.

Simon, M., and Tomlinson, G.R., 1984, Use of the Hilbert transform in modal analysis o linear and non-linear structure. *Journal of Sound and Vibration*, **96**, 421-436.

Storer, D.M., and Tomlinson, G.R., 1993, Recent developments in the measurement and interpretation of higher order transfer functions from nonlinear structures. *Mechanical Systems and Signal Processing*, **7**, 173-189.

Tang, Y.S., Boyd, S.P., and Chua, L.O., 1983, Measuring Volterra kernels. *IEEE Transactions on Circuits and Systems*, **30**, 571-577.

Tomlinson, G.R., 1980, An analysis of the distortion effects of Coulomb damping on the vector plots of lightly damped systems. *Journal of Sound and Vibration*, **71**, 443-451.

Vinh, T., Chouychai, T., Liu, H., and Djouder, M., 1987, Second order transfer function: computation and physical interpretation. In *Proceedings of the 5$^{th}$ International Mode Analysis Conference* (*IMAC*), London, UK, pp. 587-592.

Volterra, V., 1959, *Theory of Functions and of Integral and Interodifferential Equations*. New York: Dover Publication.

Wiener, N, 1958, *Nonlinear Problems in Random Theory*. New York: John Wiley.

Worden, K., Stansby, P.K., Tomlonson, G.R., and Billings, S.A., 1994, Identification of nonlinear wave forces. *Journal of Fluids and Structures*, **8**, 19-71.

Zhang, H., and Billings, S.A., 1993, Analysing nonlinear systems in the frequency domain—I: The transfer function. *Mechanical Systems and Signal Processing*, **7**, 531-550; 1994, Analysing nonlinear systems in the frequency domain—II: The phase response. *Mechanical Systems and Signal Processing*, **8**, 45-62.